\documentclass[tightenlines,twoside,secnumarabic,
               onecolumn,floatfix,nofootinbib,showpacs,11pt]{revtex4}

\usepackage[english]{babel}
\usepackage{amsmath,amssymb}
\usepackage{graphicx}
\usepackage{natbib}

\newcommand{\sprod}[2]{\ensuremath{\left\langle #1 |%
                     #2 \right\rangle}}  

\renewcommand{\vec}[1]{{\mathbf{#1}}}

\begin{document}

\pagestyle{empty}
\vspace*{4cm}
\begin{center}
\fbox{\parbox{13cm}{
{\bf NOTICE:} \\
In the first version of this article, arXiv:1005.4081v1, we had
incorrectly concluded that wave packet decoherence may be observable as
averaging of oscillations. This conclusion was based on incorrect estimates
for the size $\sigma_x$ of the neutrino wave packets---we had assumed $\sigma_x$
to be comparable to the spatial localization of the neutrino interaction
vertices, while in fact taking into account the less tight temporal localization of
the production and detection processes increases $\sigma_x$, and thus also
the coherence length $L_{\rm coh}$, by several orders of magnitude.
We are indebted to Evgeny Akhmedov, Georg Raffelt, and Leo Stodolsky
for pointing out this mistake to us.

Below, we attach the incorrect first version of the article. It will be
rewritten to explain the aforementioned problem and address related
questions.
}}
\end{center}

\clearpage


\begin{flushright}
  FERMILAB-PUB-10-168-T
\end{flushright}

\title{Testing the wave packet approach to neutrino oscillations
       in future experiments}
\author{Boris Kayser} \email[Email: ]{boris@fnal.gov}
\author{Joachim Kopp} \email[Email: ]{jkopp@fnal.gov}
\affiliation{Fermilab, Theoretical Physics Department,
             PO Box 500, Batavia, IL 60510, USA}
\date{May 25, 2010}

\begin{abstract}
  When neutrinos propagate over long distances, the mass eigenstate components
  of a flavor eigenstate will become spatially separated due to their
  different group velocities. This can happen over terrestrial distance scales
  if the neutrino energy is of order MeV and if the neutrino is localized (in a
  quantum mechanical sense) to subatomic scales. For example, if the Heisenberg
  uncertainty in the neutrino position is below $10^{-2}$~\AA, neutrino
  decoherence can be observed in reactor neutrinos using a large liquid
  scintillator detector.
\end{abstract}

\maketitle

Even though the existence of neutrino oscillations has been unambiguously
proven experimentally, the theoretical description of this phenomenon is still
occasionally disputed, see for example~\cite{Stodolsky:1998tc, Lipkin:2005kg,
Akhmedov:2008jn, Bilenky:2008ez, Akhmedov:2008zz, Ivanov:2008sd, Giunti:2008ex,
Kienert:2008nz, Cohen:2008qb, Keister:2009qn, Robertson:2010xr}.  The reason is
that neutrino oscillations, being a space- and time-dependent phenomenon,
cannot be fully described in terms of infinitely delocalized energy and
momentum eigenstates as is used in most other applications in high energy
physics. The signature dependence of neutrino oscillations on the distance
between the neutrino source and the detector obviously cannot be observed
unless the source and the detector are localized.  The uncertainty principle
then implies that the source, and consequently the neutrinos that it produces,
must be in a superposition of different momentum states. That is, the neutrino
wave function cannot be a plane wave, but must be a wave
packet~\cite{Nussinov:1976uw, Kayser:1981ye, Giunti:1991ca, Kiers:1995zj,
Grimus:1996av, Grimus:1998uh, Beuthe:2001rc, Giunti:2002xg, Giunti:2007ry,
Akhmedov:2009rb, Akhmedov:2010ms}.  For the purpose of this paper, it will be
sufficient to consider Gaussian wave packets, but our results will apply also
to more general wave packet shapes~\cite{Akhmedov:2010ms}.  We write the
neutrino wave function as
\begin{align}
  \sprod{\vec{x}}{\nu_\alpha(t)} \propto \sum_j U_{\alpha j}^*
    \exp\bigg[\!-\frac{(\vec{x} - \vec{v}_j t)^2}{4 \sigma_x^2} \bigg] \,,
\end{align}
where $\alpha$ and $j$ are flavor and mass eigenstate indices, respectively,
$U_{\alpha j}$ are the elements of the leptonic mixing matrix,
$\sigma_x$ is the width of the wave packet, which
depends on the properties of the neutrino source%
\footnote{Clearly, the observability of oscillations depends on the
localization of the detector in the same way as on that of the source. Detector
localization is incorporated into the description of oscillations
in~\cite{Giunti:2002xg,Beuthe:2001rc,Akhmedov:2010ms}},
and $\vec{v}_j$ is the group velocity corresponding to the $j$th neutrino
mass eigenstate.

As the neutrino propagates, the wave packets corresponding to different
neutrino mass eigenstates will become separated in space and time due to their
different group velocities~\cite{Nussinov:1976uw}. After a propagation distance
larger than $\sigma_x / |\vec{v}_j - \vec{v}_k|$, the wave packets of the $j$th
and $k$th mass eigenstates will have no significant overlap any more and their
coherence will be lost. Coherence can be \emph{restored} if the detection
process is delocalized over distances larger than the wave packet
separation~\cite{Kiers:1997pe}.  If this is not the case, neutrino oscillations
will be suppressed. Flavor change is still possible, but it will no longer
depend on distance.  Detailed calculations show that the $\nu_\alpha \to
\nu_\beta$ oscillation probability for ultra-relativistic neutrinos with an
average energy $E \gg m_j$ at long baseline $L$ can be written
as~\cite{Giunti:1991sx,Beuthe:2002ej}
\begin{align}
  P(\nu_\alpha \to \nu_\beta) \simeq
    \sum_{j,k} U_{\alpha j}^* U_{\alpha k} U_{\beta k}^* U_{\beta j} \,
        \exp\bigg[
          - 2\pi i \frac{L}{L^{\rm osc}_{jk}}
          - \bigg( \frac{L}{L^{\rm coh}_{jk}} \bigg)^2
        \bigg] \,,
\end{align}
with the oscillation lengths
$L^{\rm osc}_{jk} = 4\pi \bar{E} / \Delta m_{jk}^2$ and the coherence lengths
\begin{align}
  L^{\rm coh}_{jk} = \frac{4 \sqrt{2} E^2}{|\Delta m_{jk}^2|} \sigma_{x,\rm eff}
                   = 7\,347\ \text{km} \,
                     \bigg(\frac{E}{\text{MeV}}\bigg)^2
                     \bigg(\frac{7.7 \times 10^{-3}\text{eV}^2}{\Delta m_{jk}^2}\bigg)
                     \bigg(\frac{\sigma_{x,\rm eff}}{\text{\AA}}\bigg) \,.
  \label{eq:Lcoh}
\end{align}
Here $\sigma_{x,\rm eff}$ is an effective wave packet width that depends
on the spatial delocalization of the source and that of the detector, and is
dominated by the larger of the two.%
\footnote{$\sigma_{x,\rm eff}$ depends also on the temporal delocalization
of the production and detection processes, which, however, is usually similar
to the spatial delocalization.}

Eq.~\eqref{eq:Lcoh} shows that neutrinos from astrophysical sources always
arrive at the Earth as completely incoherent mixtures of mass eigenstates,
while observation of decoherence effects in terrestrial experiments would
require a long baseline, low neutrino energy, and small $\sigma_{x,\rm eff}$.

These requirements can be fulfilled for reactor neutrinos observed in a future
large liquid scintillator detector like Hanohano~\cite{Learned:2008zj} or
LENA~\cite{Wurm:2010ny}.  The reactor neutrino event rate peaks at $E
\sim 4$~MeV, and the distance from a multipurpose detector like Hanohano or LENA
to the nearest nuclear power station will be of order $L \gtrsim
100$~km to avoid large backgrounds in geo-neutrino, supernova relic neutrino,
and proton decay studies.

It is more difficult to estimate the wave packet width $\sigma_{x,\rm eff}$
entering in eq.~\eqref{eq:Lcoh}. Since it depends on the quantum mechanical
localization of the neutrino in space, we have to ask how well we can \emph{in
principle} determine the position of the neutrino production and detection
points, given perfect experimental equipment.

For neutrinos emitted from free particles in flight, the spatial uncertainty
of the production process will be similar to the mean free path of the
parent particle~\cite{Torres:2007zz}.

A neutrino emission or detection process in a solid or liquid is localized at
least to interatomic distance scales of $\sigma_{x,\rm eff} \sim
\mathcal{O}(1-10\ \text{\AA})$ because the emitting or absorbing atom is
continuously interacting with its neighbors.  The latter can be viewed as a
thermal bath in the sense that their quantum states at different times are
completely uncorrelated. An interaction of a particle with a thermal bath
constitutes a measurement in the quantum mechanical sense that localizes
the particle.

Another reason why $\sigma_{x,\rm eff}$ cannot be larger than an interatomic
distance is that we can in principle measure the location of the production and
detection vertices to that accuracy in an experiment.  Whenever we detect that
a neutrino has been produced or absorbed (for example by detecting the
associated charged lepton), we can in principle quickly bombard the source with
a high-energy beam of probe particles to determine which nucleus has undergone
a transition in the process.

Finally, a spatial uncertainty much larger than a few \AA\ would imply that
\emph{all} particles participating in the process would have to be delocalized
over many interatomic distances.  Imagine an energetic charged particle like an
outgoing charged lepton traveling through a solid or liquid material and
interacting with the atoms.  For concreteness, let us assume that it excites
the atomic shells, which subsequently relax by emitting scintillation light.
If the energetic particle was delocalized over many interatomic distances, it
would be impossible to tell which atoms is excited.  The resulting
scintillation light would then behave as if it came from several atoms
simultaneously and would therefore exhibit interference patterns. We are not
aware of any experimental evidence for such interference patterns.

For neutrino emission or detection in a solid state crystal, the upper limit on
the delocalization scale can be even tighter than 1--10~\AA. To be specific,
consider neutrino production in a reactor fuel rod.  Once we know the position
of the production vertex to within one lattice spacing---which is possible
according to the above arguments---we can make use of our knowledge of the
crystal structure to pin down the exact lattice site where the process must
have occurred.  In a crystal at zero temperature, the emitting nucleus would be
localized at its lattice site to within its own size of order $\text{few}
\times 10^{-15}$~m. At finite temperature $T$, it will oscillate about this
position, with the amplitude of these vibrations being of order $[T / m \,
\Theta_D^2]^{1/2} \sim$~0.01--0.1~\AA\ for Debye temperatures $\Theta_D$ of few
hundred Kelvin and nuclear masses $m \sim 100$~GeV.  To arrive at this estimate,
we have assumed the nucleus to be bound in a harmonic oscillator potential $V =
\tfrac{1}{2} m \omega^2 x^2$ with characteristic frequency $\omega = \Theta_D$,
and the thermal excitation energy to be of order $T$. We conclude that the
spatial uncertainty associated with a neutrino production process in a solid
state source is roughly between $10^{-5}$~\AA\ and 0.1~\AA, depending on the
temperature. Similar arguments apply to solid state neutrino detectors.

At least for reactor neutrino experiments with liquid scintillator detectors,
we can also set a \emph{lower} limit on $\sigma_{x,\rm eff}$ from the fact that the
KamLAND experiment has observed neutrino oscillation consistent with the
results from solar neutrino experiments at baselines of $\mathcal{O}(100\
\textrm{km})$~\cite{Eguchi:2002dm}.%
\footnote{The possibility cannot be ruled out that KamLAND is affected by neutrino
decoherence at a subdominant level. In this case, a reanalysis would be required,
leading to modifications of the global fit of the solar oscillation parameters.
The experiment we propose in this paper would be able to determine if such a
reanalysis is indeed necessary. Decoherence effects in KamLAND have been studied
for example in~\cite{Araki:2004mb,Fogli:2007tx}.}
This translates into the limit $\sigma_{x,\rm eff} \gtrsim 10^{-3}$~\AA.

From the above considerations we conclude that, for reactor neutrino
experiments
\begin{align}
  10^{-3}\ \text{\AA} \lesssim \sigma_{x,\rm eff} \lesssim 10^1\ \text{\AA}
  \label{eq:decoh-limits}
\end{align}
or
\begin{align}
  100\ \text{km} \lesssim L^{\rm coh}_{21} \lesssim 1\,000\,000\ \text{km}
\end{align}
These estimates show that it may well be possible to detect neutrino wave
packet decoherence effects in terrestrial experiments; this would prove
that the wave packet formalism is indeed well-suited to describe neutrino
oscillations, and a measurement of the coherence length would provide
interesting information about the quantum mechanics of the production and
detection processes.

The experimental prospects for detecting neutrino decoherence have been studied
previously in the context of atmospheric neutrinos~\cite{Lisi:2000zt},
short-baseline ($\mathcal{O}({\rm km})$) reactor
experiments~\cite{Blennow:2005yk}, superbeams~\cite{Ribeiro:2007jq,
Sakharov:2009rn}, and a neutrino factory~\cite{Blennow:2005yk}. Here, we will
study the prospects of observing the predicted wave packet decoherence from
eq.~\eqref{eq:decoh-limits} by observing reactor neutrinos at very long
baselines of at least a few hundred km in a large liquid scintillator detector
like Hanohano or LENA.  We have simulated this setup using
GLoBES~\cite{Huber:2004ka,Huber:2007ji}, assuming a 5-year run of a
LENA-like~\cite{Wurm:2010ny} detector with a fiducial mass of 45~kt and a
Gaussian energy resolution of $0.07 \times (E/\text{MeV} - 0.8)^{1/2}$~MeV.  As
systematic uncertainties, we include a 3\% error on the reactor neutrino
flux, a 50\% uncertainty on the geo-neutrino flux, a 0.5\% energy calibration
error, and a 0.5\% spectral error, uncorrelated between different energy bins.
We study three different detector sites: The Pyh\"asalmi mine in Finland, the
Deep Underground Science and Engineering Laboratory (DUSEL) in South Dakota,
USA, and a site in Hawaii. Unless otherwise noted, we take into account the
neutrino flux from all nuclear power stations in the world~\cite{IAEAreactors},
as well as the geo-neutrino background, which dominates the event rates below
$E \lesssim 3.3$~MeV.%
\footnote{Note that the data on reactor sites available to
us dates from 2000, so some recently commissioned stations may not be
included. To partly compensate for this, we assume all stations that were under
construction in 2000 to be operational at full power by now.}
The reactor neutrino spectrum we are using is based on~\cite{Murayama:2000iq,
Eguchi:2002dm}, while the geo-neutrino spectrum has been kindly provided to us
in machine-readable form by the author of ref.~\cite{Enomoto:2006}. The cross
sections for neutrino detection in inverse beta decay are from
ref.~\cite{Vogel:1999zy}.

\begin{figure}
  \begin{center}
    \includegraphics[width=\textwidth]{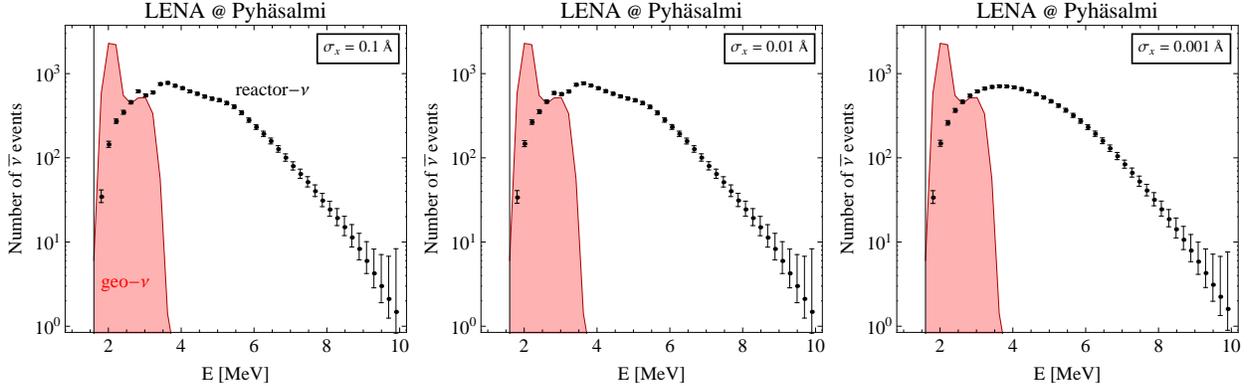}
  \end{center}
  \vspace{-0.5cm}
  \caption{The neutrino spectrum in a 45~kt (fiducial) liquid scintillator
    detector located in the Pyh\"asalmi mine in Finland for different
    values of the wave packet width $\sigma_{x,\rm eff}$. The red curve
    is the geo-neutrino background, while the black dots are the reactor
    neutrino signal.}
  \label{fig:spectra}
\end{figure}

In fig.~\ref{fig:spectra}, we show the effect of decoherence on the neutrino
spectrum expected at the Pyh\"asalmi mine in Finland.  For large $\sigma_{x,\rm
eff}$, where $L^{\rm coh} \gg L$, an oscillation pattern is visible, even
though it is smeared due to the overlap of signals from different reactors
at various baselines. For small $\sigma_{x,\rm eff}$ oscillations are
completely smoothed out by decoherence. 

\begin{figure}
  \begin{center}
    \includegraphics[width=8cm]{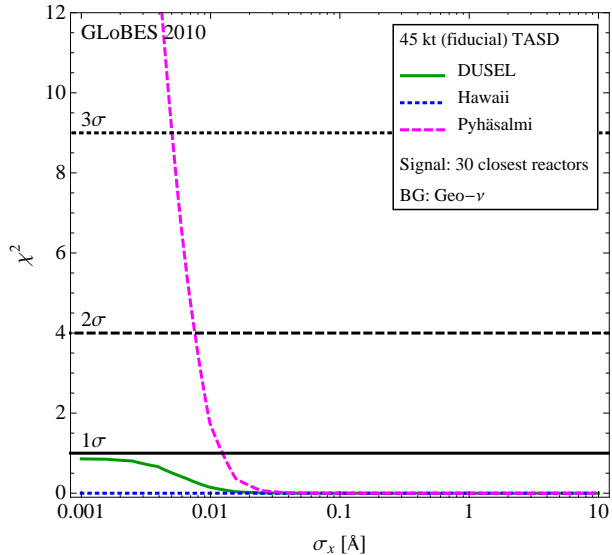}
  \end{center}
  \vspace{-0.5cm}
  \caption{Discovery reach for neutrino decoherence in a 45~kt (fiducial)
    detector located at DUSEL, Hawaii, or Pyh\"asalmi. Only the Pyh\"asalmi
    site allows for good sensitivity to decoherence effects.}
  \label{fig:chi2}
\end{figure}

To quantify this observation and determine the discovery potential for
decoherence effects, we have simulated the expected event spectrum for
different values of $\sigma_{x,\rm eff}$, and have then performed a fit
assuming no decoherence. The resulting $\chi^2$ is plotted in
fig.~\ref{fig:chi2}. We find that prospects for observing decoherence are best
at the Pyh\"asalmi site due to the proximity of only few nuclear reactors. This
keeps washout of oscillations due to the superposition of many reactor spectra
at many different baselines small. On the other hand, there is no nearby
reactor that would make the detector blind to events from distant reactors that
could carry information about decoherence.  In the DUSEL scenario, there are no
nearby reactors either, but washout due to the larger number of nuclear
reactors in the eastern United States is a problem.  In Hawaii, on the other
hand, the nearest nuclear reactors are thousands of kilometers away, so that
the number of events is too small for the presence or absence of an oscillation
pattern to be seen. Moreover, at such long baselines, consecutive oscillation
maxima are very close in energy, so the detector resolution becomes an issue.
Thus we conclude that the optimal detector site for a decoherence measurement
is one that has a few (but not too many) nuclear reactors at baselines of
several 100~km.

\begin{figure}
  \begin{center}
    \includegraphics{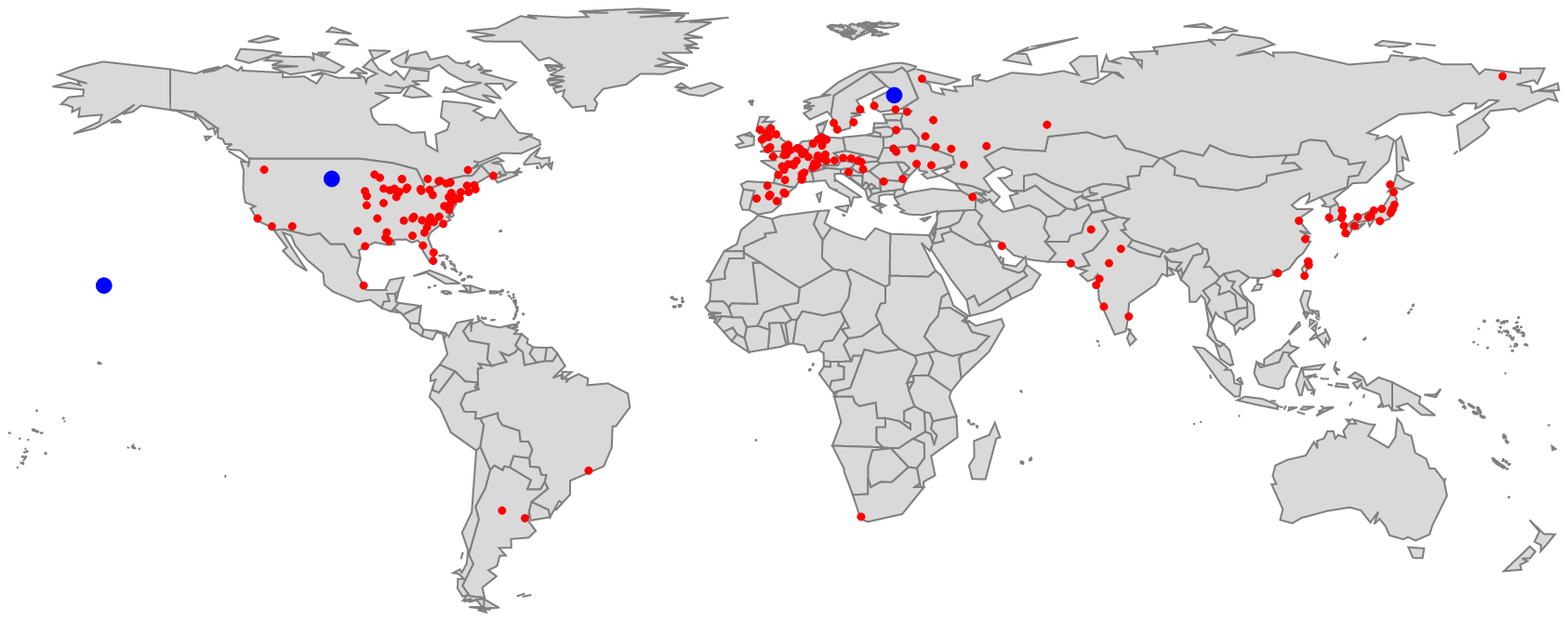}
    \includegraphics{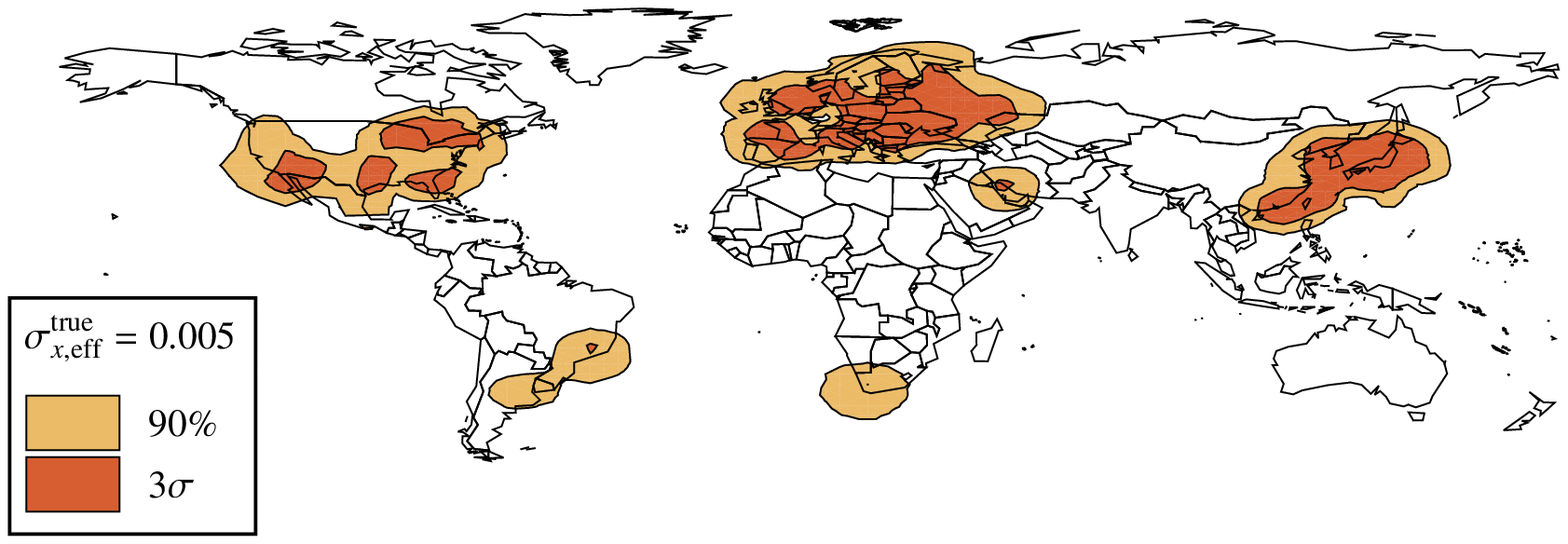}
  \end{center}
  \vspace{-1.0cm}
  \caption{Top: A map of all nuclear reactors running or under construction as
    of 2000 (red dots), together with the proposed detector sites at Hawaii,
    Pyh\"asalmi, and DUSEL (blue dots). Data on nuclear reactors is
    from~\cite{IAEAreactors}.  Bottom: Discovery reach for wave packet
    decoherence with $\sigma_{x,\rm eff} = 0.005$~\AA\ at different
    geographic locations. Note that, due to limited resolution, the plot hides
    the fact that very close to any nuclear reactor, the sensitivity greatly
    decreases because the detector is blinded by short-baseline neutrinos in that
    case.}
  \label{fig:map}
\end{figure}

Fig.~\ref{fig:map} shows that this condition is fulfilled in many locations
around the world. In the top panel, we have mapped the locations of nuclear
power stations~\cite{IAEAreactors} (as of 2000), while the bottom panel shows
the significance at which wave packet decoherence can be detected if the wave
packet width is 0.005~\AA, close to the lower end of the allowed range from
eq.~\eqref{eq:decoh-limits}.  We have checked that a larger values
$\sigma_{x,\rm eff} = 0.01$ can only be detected in few places in Japan, while
even larger values cannot be detected anywhere in the world with the
experimental setup we have simulated. Note that for each grid point in
fig.~\ref{fig:map}, we have simulated only the 32 closest nuclear reactors
to speed up the numerical computations.

In conclusion, we have presented theoretical arguments showing that neutrino
wave packets in a reactor experiment should have a width between 0.001~\AA\ and
10~\AA. The lower limit is based on the observation of oscillations at KamLAND,
while the upper limit has been estimated by considering the localization of
typical neutrino production and detection processes. These numbers imply that
wave packet decoherence due to different group velocities of different neutrino
mass eigenstates can occur over distances of $10^2$--$10^6$~km for
$\mathcal{O}(\text{few } \textrm{MeV})$ neutrinos.  In higher energy experiments
or in experiments in which neutrinos are produced by decay in flight of some
parent particle, the wave packet width can be much larger, while in a
hypothetical experiment in which $\mathcal{O}(\text{few } \textrm{MeV})$
neutrinos are emitted and detected by nuclei bound in solid state lattices at
very low temperature, the localization can be of $\mathcal{O}(\text{few} \times
10^{-5})$~\AA, corresponding to a coherence length of order 1~km. We have then
performed detailed numerical simulation to show that wave packet decoherence is
observable in a future large liquid scintillator detector like Hanohano or LENA,
but only if the coherence length is at the lower end of the expected
range and if the detector site is far (few hundred kilometers), but not too far
($\lesssim 1\,000$~km) from the nearest nuclear power station.

The authors are indebted to Walter Potzel and Michael Wurm for very useful
discussions on large liquid scintillator detectors. Fermilab is operated by Fermi
Research Alliance, LLC under Contract No.~DE-AC02-07CH11359 with the US
Department of Energy.



\begin{thebibliography}{41}
\expandafter\ifx\csname natexlab\endcsname\relax\def\natexlab#1{#1}\fi
\expandafter\ifx\csname bibnamefont\endcsname\relax
  \def\bibnamefont#1{#1}\fi
\expandafter\ifx\csname bibfnamefont\endcsname\relax
  \def\bibfnamefont#1{#1}\fi
\expandafter\ifx\csname citenamefont\endcsname\relax
  \def\citenamefont#1{#1}\fi
\expandafter\ifx\csname url\endcsname\relax
  \def\url#1{\texttt{#1}}\fi
\expandafter\ifx\csname urlprefix\endcsname\relax\def\urlprefix{URL }\fi
\providecommand{\bibinfo}[2]{#2}
\providecommand{\eprint}[2][]{\url{#2}}

\bibitem[{\citenamefont{Stodolsky}(1998)}]{Stodolsky:1998tc}
\bibinfo{author}{\bibfnamefont{L.}~\bibnamefont{Stodolsky}},
  \bibinfo{journal}{Phys. Rev.} \textbf{\bibinfo{volume}{D58}},
  \bibinfo{pages}{036006} (\bibinfo{year}{1998}), \eprint{hep-ph/9802387}.

\bibitem[{\citenamefont{Lipkin}(2006)}]{Lipkin:2005kg}
\bibinfo{author}{\bibfnamefont{H.~J.} \bibnamefont{Lipkin}},
  \bibinfo{journal}{Phys.Lett.} \textbf{\bibinfo{volume}{B642}},
  \bibinfo{pages}{366} (\bibinfo{year}{2006}), \eprint{hep-ph/0505141}.

\bibitem[{\citenamefont{Akhmedov
  et~al.}(2008{\natexlab{a}})\citenamefont{Akhmedov, Kopp, and
  Lindner}}]{Akhmedov:2008jn}
\bibinfo{author}{\bibfnamefont{E.~K.} \bibnamefont{Akhmedov}},
  \bibinfo{author}{\bibfnamefont{J.}~\bibnamefont{Kopp}}, \bibnamefont{and}
  \bibinfo{author}{\bibfnamefont{M.}~\bibnamefont{Lindner}},
  \bibinfo{journal}{JHEP} \textbf{\bibinfo{volume}{05}}, \bibinfo{pages}{005}
  (\bibinfo{year}{2008}{\natexlab{a}}), \eprint{0802.2513}.

\bibitem[{\citenamefont{Bilenky et~al.}(2008)\citenamefont{Bilenky, von
  Feilitzsch, and Potzel}}]{Bilenky:2008ez}
\bibinfo{author}{\bibfnamefont{S.~M.} \bibnamefont{Bilenky}},
  \bibinfo{author}{\bibfnamefont{F.}~\bibnamefont{von Feilitzsch}},
  \bibnamefont{and} \bibinfo{author}{\bibfnamefont{W.}~\bibnamefont{Potzel}},
  \bibinfo{journal}{J. Phys.} \textbf{\bibinfo{volume}{G35}},
  \bibinfo{pages}{095003} (\bibinfo{year}{2008}), \eprint{0803.0527}.

\bibitem[{\citenamefont{Akhmedov
  et~al.}(2008{\natexlab{b}})\citenamefont{Akhmedov, Kopp, and
  Lindner}}]{Akhmedov:2008zz}
\bibinfo{author}{\bibfnamefont{E.~K.} \bibnamefont{Akhmedov}},
  \bibinfo{author}{\bibfnamefont{J.}~\bibnamefont{Kopp}}, \bibnamefont{and}
  \bibinfo{author}{\bibfnamefont{M.}~\bibnamefont{Lindner}}
  (\bibinfo{year}{2008}{\natexlab{b}}), \eprint{0803.1424}.

\bibitem[{\citenamefont{Ivanov et~al.}(2008)\citenamefont{Ivanov, Reda, and
  Kienle}}]{Ivanov:2008sd}
\bibinfo{author}{\bibfnamefont{A.~N.} \bibnamefont{Ivanov}},
  \bibinfo{author}{\bibfnamefont{R.}~\bibnamefont{Reda}}, \bibnamefont{and}
  \bibinfo{author}{\bibfnamefont{P.}~\bibnamefont{Kienle}}
  (\bibinfo{year}{2008}), \eprint{arXiv:0801.2121 [nucl-th]}.

\bibitem[{\citenamefont{Giunti}(2008)}]{Giunti:2008ex}
\bibinfo{author}{\bibfnamefont{C.}~\bibnamefont{Giunti}}
  (\bibinfo{year}{2008}), \eprint{arXiv:0801.4639 [hep-ph]}.

\bibitem[{\citenamefont{Kienert et~al.}(2008)\citenamefont{Kienert, Kopp,
  Lindner, and Merle}}]{Kienert:2008nz}
\bibinfo{author}{\bibfnamefont{H.}~\bibnamefont{Kienert}},
  \bibinfo{author}{\bibfnamefont{J.}~\bibnamefont{Kopp}},
  \bibinfo{author}{\bibfnamefont{M.}~\bibnamefont{Lindner}}, \bibnamefont{and}
  \bibinfo{author}{\bibfnamefont{A.}~\bibnamefont{Merle}}
  (\bibinfo{year}{2008}), \eprint{0808.2389}.

\bibitem[{\citenamefont{Cohen et~al.}(2008)\citenamefont{Cohen, Glashow, and
  Ligeti}}]{Cohen:2008qb}
\bibinfo{author}{\bibfnamefont{A.~G.} \bibnamefont{Cohen}},
  \bibinfo{author}{\bibfnamefont{S.~L.} \bibnamefont{Glashow}},
  \bibnamefont{and} \bibinfo{author}{\bibfnamefont{Z.}~\bibnamefont{Ligeti}}
  (\bibinfo{year}{2008}), \eprint{0810.4602}.

\bibitem[{\citenamefont{Keister and Polyzou}(2010)}]{Keister:2009qn}
\bibinfo{author}{\bibfnamefont{B.}~\bibnamefont{Keister}} \bibnamefont{and}
  \bibinfo{author}{\bibfnamefont{W.}~\bibnamefont{Polyzou}},
  \bibinfo{journal}{Phys.Scripta} \textbf{\bibinfo{volume}{81}},
  \bibinfo{pages}{055102} (\bibinfo{year}{2010}), \eprint{arXiv:0908.1404}.

\bibitem[{\citenamefont{Robertson}(2010)}]{Robertson:2010xr}
\bibinfo{author}{\bibfnamefont{R.~G.~H.} \bibnamefont{Robertson}}
  (\bibinfo{year}{2010}), \eprint{1004.1847}.

\bibitem[{\citenamefont{Nussinov}(1976)}]{Nussinov:1976uw}
\bibinfo{author}{\bibfnamefont{S.}~\bibnamefont{Nussinov}},
  \bibinfo{journal}{Phys. Lett.} \textbf{\bibinfo{volume}{B63}},
  \bibinfo{pages}{201} (\bibinfo{year}{1976}).

\bibitem[{\citenamefont{Kayser}(1981)}]{Kayser:1981ye}
\bibinfo{author}{\bibfnamefont{B.}~\bibnamefont{Kayser}},
  \bibinfo{journal}{Phys. Rev.} \textbf{\bibinfo{volume}{D24}},
  \bibinfo{pages}{110} (\bibinfo{year}{1981}).

\bibitem[{\citenamefont{Giunti et~al.}(1991)\citenamefont{Giunti, Kim, and
  Lee}}]{Giunti:1991ca}
\bibinfo{author}{\bibfnamefont{C.}~\bibnamefont{Giunti}},
  \bibinfo{author}{\bibfnamefont{C.~W.} \bibnamefont{Kim}}, \bibnamefont{and}
  \bibinfo{author}{\bibfnamefont{U.~W.} \bibnamefont{Lee}},
  \bibinfo{journal}{Phys. Rev.} \textbf{\bibinfo{volume}{D44}},
  \bibinfo{pages}{3635} (\bibinfo{year}{1991}).

\bibitem[{\citenamefont{Kiers et~al.}(1996)\citenamefont{Kiers, Nussinov, and
  Weiss}}]{Kiers:1995zj}
\bibinfo{author}{\bibfnamefont{K.}~\bibnamefont{Kiers}},
  \bibinfo{author}{\bibfnamefont{S.}~\bibnamefont{Nussinov}}, \bibnamefont{and}
  \bibinfo{author}{\bibfnamefont{N.}~\bibnamefont{Weiss}},
  \bibinfo{journal}{Phys. Rev.} \textbf{\bibinfo{volume}{D53}},
  \bibinfo{pages}{537} (\bibinfo{year}{1996}), \eprint{hep-ph/9506271}.

\bibitem[{\citenamefont{Grimus and Stockinger}(1996)}]{Grimus:1996av}
\bibinfo{author}{\bibfnamefont{W.}~\bibnamefont{Grimus}} \bibnamefont{and}
  \bibinfo{author}{\bibfnamefont{P.}~\bibnamefont{Stockinger}},
  \bibinfo{journal}{Phys. Rev.} \textbf{\bibinfo{volume}{D54}},
  \bibinfo{pages}{3414} (\bibinfo{year}{1996}), \eprint{hep-ph/9603430}.

\bibitem[{\citenamefont{Grimus et~al.}(1999)\citenamefont{Grimus, Stockinger,
  and Mohanty}}]{Grimus:1998uh}
\bibinfo{author}{\bibfnamefont{W.}~\bibnamefont{Grimus}},
  \bibinfo{author}{\bibfnamefont{P.}~\bibnamefont{Stockinger}},
  \bibnamefont{and} \bibinfo{author}{\bibfnamefont{S.}~\bibnamefont{Mohanty}},
  \bibinfo{journal}{Phys. Rev.} \textbf{\bibinfo{volume}{D59}},
  \bibinfo{pages}{013011} (\bibinfo{year}{1999}), \eprint{hep-ph/9807442}.

\bibitem[{\citenamefont{Beuthe}(2003)}]{Beuthe:2001rc}
\bibinfo{author}{\bibfnamefont{M.}~\bibnamefont{Beuthe}},
  \bibinfo{journal}{Phys. Rept.} \textbf{\bibinfo{volume}{375}},
  \bibinfo{pages}{105} (\bibinfo{year}{2003}), \eprint{hep-ph/0109119}.

\bibitem[{\citenamefont{Giunti}(2002)}]{Giunti:2002xg}
\bibinfo{author}{\bibfnamefont{C.}~\bibnamefont{Giunti}},
  \bibinfo{journal}{JHEP} \textbf{\bibinfo{volume}{11}}, \bibinfo{pages}{017}
  (\bibinfo{year}{2002}), \eprint{hep-ph/0205014}.

\bibitem[{\citenamefont{Giunti and Kim}(2007)}]{Giunti:2007ry}
\bibinfo{author}{\bibfnamefont{C.}~\bibnamefont{Giunti}} \bibnamefont{and}
  \bibinfo{author}{\bibfnamefont{C.~W.} \bibnamefont{Kim}},
  \emph{\bibinfo{title}{{Fundamentals of Neutrino Physics and Astrophysics}}}
  (\bibinfo{publisher}{Oxford University Press}, \bibinfo{address}{Oxford, UK},
  \bibinfo{year}{2007}).

\bibitem[{\citenamefont{Akhmedov and Smirnov}(2009)}]{Akhmedov:2009rb}
\bibinfo{author}{\bibfnamefont{E.~K.} \bibnamefont{Akhmedov}} \bibnamefont{and}
  \bibinfo{author}{\bibfnamefont{A.~Y.} \bibnamefont{Smirnov}}
  (\bibinfo{year}{2009}), \eprint{0905.1903}.

\bibitem[{\citenamefont{Akhmedov and Kopp}(2010)}]{Akhmedov:2010ms}
\bibinfo{author}{\bibfnamefont{E.~K.} \bibnamefont{Akhmedov}} \bibnamefont{and}
  \bibinfo{author}{\bibfnamefont{J.}~\bibnamefont{Kopp}},
  \bibinfo{journal}{JHEP} \textbf{\bibinfo{volume}{04}}, \bibinfo{pages}{008}
  (\bibinfo{year}{2010}), \eprint{1001.4815}.

\bibitem[{\citenamefont{Kiers and Weiss}(1998)}]{Kiers:1997pe}
\bibinfo{author}{\bibfnamefont{K.}~\bibnamefont{Kiers}} \bibnamefont{and}
  \bibinfo{author}{\bibfnamefont{N.}~\bibnamefont{Weiss}},
  \bibinfo{journal}{Phys. Rev.} \textbf{\bibinfo{volume}{D57}},
  \bibinfo{pages}{3091} (\bibinfo{year}{1998}), \eprint{hep-ph/9710289}.

\bibitem[{\citenamefont{Giunti et~al.}(1992)\citenamefont{Giunti, Kim, and
  Lee}}]{Giunti:1991sx}
\bibinfo{author}{\bibfnamefont{C.}~\bibnamefont{Giunti}},
  \bibinfo{author}{\bibfnamefont{C.~W.} \bibnamefont{Kim}}, \bibnamefont{and}
  \bibinfo{author}{\bibfnamefont{U.~W.} \bibnamefont{Lee}},
  \bibinfo{journal}{Phys. Lett.} \textbf{\bibinfo{volume}{B274}},
  \bibinfo{pages}{87} (\bibinfo{year}{1992}).

\bibitem[{\citenamefont{Beuthe}(2002)}]{Beuthe:2002ej}
\bibinfo{author}{\bibfnamefont{M.}~\bibnamefont{Beuthe}},
  \bibinfo{journal}{Phys. Rev.} \textbf{\bibinfo{volume}{D66}},
  \bibinfo{pages}{013003} (\bibinfo{year}{2002}), \eprint{hep-ph/0202068}.

\bibitem[{\citenamefont{Learned et~al.}(2008)\citenamefont{Learned, Dye, and
  Pakvasa}}]{Learned:2008zj}
\bibinfo{author}{\bibfnamefont{J.~G.} \bibnamefont{Learned}},
  \bibinfo{author}{\bibfnamefont{S.~T.} \bibnamefont{Dye}}, \bibnamefont{and}
  \bibinfo{author}{\bibfnamefont{S.}~\bibnamefont{Pakvasa}}
  (\bibinfo{year}{2008}), \eprint{0810.4975}.

\bibitem[{\citenamefont{Wurm et~al.}(2010)}]{Wurm:2010ny}
\bibinfo{author}{\bibfnamefont{M.}~\bibnamefont{Wurm}} \bibnamefont{et~al.}
  (\bibinfo{year}{2010}), \eprint{1004.3474}.

\bibitem[{\citenamefont{Torres and Guzzo}(2007)}]{Torres:2007zz}
\bibinfo{author}{\bibfnamefont{F.}~\bibnamefont{Torres}} \bibnamefont{and}
  \bibinfo{author}{\bibfnamefont{M.}~\bibnamefont{Guzzo}},
  \bibinfo{journal}{Braz.J.Phys.} \textbf{\bibinfo{volume}{37}},
  \bibinfo{pages}{1273} (\bibinfo{year}{2007}).

\bibitem[{\citenamefont{Eguchi et~al.}(2003)}]{Eguchi:2002dm}
\bibinfo{author}{\bibfnamefont{K.}~\bibnamefont{Eguchi}} \bibnamefont{et~al.}
  (\bibinfo{collaboration}{KamLAND}), \bibinfo{journal}{Phys. Rev. Lett.}
  \textbf{\bibinfo{volume}{90}}, \bibinfo{pages}{021802}
  (\bibinfo{year}{2003}), \eprint{hep-ex/0212021}.

\bibitem[{\citenamefont{Araki et~al.}(2005)}]{Araki:2004mb}
\bibinfo{author}{\bibfnamefont{T.}~\bibnamefont{Araki}} \bibnamefont{et~al.}
  (\bibinfo{collaboration}{KamLAND}), \bibinfo{journal}{Phys. Rev. Lett.}
  \textbf{\bibinfo{volume}{94}}, \bibinfo{pages}{081801}
  (\bibinfo{year}{2005}), \eprint{hep-ex/0406035}.

\bibitem[{\citenamefont{Fogli et~al.}(2007)\citenamefont{Fogli, Lisi, Marrone,
  Montanino, and Palazzo}}]{Fogli:2007tx}
\bibinfo{author}{\bibfnamefont{G.}~\bibnamefont{Fogli}},
  \bibinfo{author}{\bibfnamefont{E.}~\bibnamefont{Lisi}},
  \bibinfo{author}{\bibfnamefont{A.}~\bibnamefont{Marrone}},
  \bibinfo{author}{\bibfnamefont{D.}~\bibnamefont{Montanino}},
  \bibnamefont{and} \bibinfo{author}{\bibfnamefont{A.}~\bibnamefont{Palazzo}},
  \bibinfo{journal}{Phys.Rev.} \textbf{\bibinfo{volume}{D76}},
  \bibinfo{pages}{033006} (\bibinfo{year}{2007}), \eprint{arXiv:0704.2568}.

\bibitem[{\citenamefont{Lisi et~al.}(2000)\citenamefont{Lisi, Marrone, and
  Montanino}}]{Lisi:2000zt}
\bibinfo{author}{\bibfnamefont{E.}~\bibnamefont{Lisi}},
  \bibinfo{author}{\bibfnamefont{A.}~\bibnamefont{Marrone}}, \bibnamefont{and}
  \bibinfo{author}{\bibfnamefont{D.}~\bibnamefont{Montanino}},
  \bibinfo{journal}{Phys. Rev. Lett.} \textbf{\bibinfo{volume}{85}},
  \bibinfo{pages}{1166} (\bibinfo{year}{2000}), \eprint{hep-ph/0002053}.

\bibitem[{\citenamefont{Blennow et~al.}(2005)\citenamefont{Blennow, Ohlsson,
  and Winter}}]{Blennow:2005yk}
\bibinfo{author}{\bibfnamefont{M.}~\bibnamefont{Blennow}},
  \bibinfo{author}{\bibfnamefont{T.}~\bibnamefont{Ohlsson}}, \bibnamefont{and}
  \bibinfo{author}{\bibfnamefont{W.}~\bibnamefont{Winter}},
  \bibinfo{journal}{JHEP} \textbf{\bibinfo{volume}{0506}}, \bibinfo{pages}{049}
  (\bibinfo{year}{2005}), \eprint{hep-ph/0502147}.

\bibitem[{\citenamefont{Ribeiro et~al.}(2008)}]{Ribeiro:2007jq}
\bibinfo{author}{\bibfnamefont{N.~C.} \bibnamefont{Ribeiro}}
  \bibnamefont{et~al.}, \bibinfo{journal}{Phys. Rev.}
  \textbf{\bibinfo{volume}{D77}}, \bibinfo{pages}{073007}
  (\bibinfo{year}{2008}), \eprint{0712.4314}.

\bibitem[{\citenamefont{Sakharov et~al.}(2009)\citenamefont{Sakharov,
  Mavromatos, Meregaglia, Rubbia, and Sarkar}}]{Sakharov:2009rn}
\bibinfo{author}{\bibfnamefont{A.}~\bibnamefont{Sakharov}},
  \bibinfo{author}{\bibfnamefont{N.}~\bibnamefont{Mavromatos}},
  \bibinfo{author}{\bibfnamefont{A.}~\bibnamefont{Meregaglia}},
  \bibinfo{author}{\bibfnamefont{A.}~\bibnamefont{Rubbia}}, \bibnamefont{and}
  \bibinfo{author}{\bibfnamefont{S.}~\bibnamefont{Sarkar}}
  (\bibinfo{year}{2009}), \eprint{0903.4985}.

\bibitem[{\citenamefont{Huber et~al.}(2005)\citenamefont{Huber, Lindner, and
  Winter}}]{Huber:2004ka}
\bibinfo{author}{\bibfnamefont{P.}~\bibnamefont{Huber}},
  \bibinfo{author}{\bibfnamefont{M.}~\bibnamefont{Lindner}}, \bibnamefont{and}
  \bibinfo{author}{\bibfnamefont{W.}~\bibnamefont{Winter}},
  \bibinfo{journal}{Comput. Phys. Commun.} \textbf{\bibinfo{volume}{167}},
  \bibinfo{pages}{195} (\bibinfo{year}{2005}), \eprint{hep-ph/0407333},
  \urlprefix\url{http://www.mpi-hd.mpg.de/~globes}.

\bibitem[{\citenamefont{Huber et~al.}(2007)\citenamefont{Huber, Kopp, Lindner,
  Rolinec, and Winter}}]{Huber:2007ji}
\bibinfo{author}{\bibfnamefont{P.}~\bibnamefont{Huber}},
  \bibinfo{author}{\bibfnamefont{J.}~\bibnamefont{Kopp}},
  \bibinfo{author}{\bibfnamefont{M.}~\bibnamefont{Lindner}},
  \bibinfo{author}{\bibfnamefont{M.}~\bibnamefont{Rolinec}}, \bibnamefont{and}
  \bibinfo{author}{\bibfnamefont{W.}~\bibnamefont{Winter}},
  \bibinfo{journal}{Comput. Phys. Commun.} \textbf{\bibinfo{volume}{177}},
  \bibinfo{pages}{432} (\bibinfo{year}{2007}), \eprint{hep-ph/0701187},
  \urlprefix\url{http://www.mpi-hd.mpg.de/~globes}.

\bibitem[{\citenamefont{{The United Nations Environment
  Programme}}(1999)}]{IAEAreactors}
\bibinfo{author}{\bibnamefont{{The United Nations Environment Programme}}},
  \emph{\bibinfo{title}{Nuclear power stations of the world}}
  (\bibinfo{year}{1999}),
  \urlprefix\url{http://www.grid.unep.ch/GRID_search_details.php?dataid=GNV181%
}.

\bibitem[{\citenamefont{Murayama and Pierce}(2002)}]{Murayama:2000iq}
\bibinfo{author}{\bibfnamefont{H.}~\bibnamefont{Murayama}} \bibnamefont{and}
  \bibinfo{author}{\bibfnamefont{A.}~\bibnamefont{Pierce}},
  \bibinfo{journal}{Phys. Rev.} \textbf{\bibinfo{volume}{D65}},
  \bibinfo{pages}{013012} (\bibinfo{year}{2002}), \eprint{hep-ph/0012075}.

\bibitem[{\citenamefont{Enomoto}(2005)}]{Enomoto:2006}
\bibinfo{author}{\bibfnamefont{S.}~\bibnamefont{Enomoto}}, Ph.D. thesis,
  \bibinfo{school}{Tohoku University} (\bibinfo{year}{2005}),
  \bibinfo{note}{available from
  http://kamland.stanford.edu/GeoNeutrinos/GeoNuResult/}.

\bibitem[{\citenamefont{Vogel and Beacom}(1999)}]{Vogel:1999zy}
\bibinfo{author}{\bibfnamefont{P.}~\bibnamefont{Vogel}} \bibnamefont{and}
  \bibinfo{author}{\bibfnamefont{J.~F.} \bibnamefont{Beacom}},
  \bibinfo{journal}{Phys. Rev.} \textbf{\bibinfo{volume}{D60}},
  \bibinfo{pages}{053003} (\bibinfo{year}{1999}), \eprint{hep-ph/9903554}.

\end{thebibliography}
\end{document}